\documentclass[twoside,twocolumn,9pt]{article}
\usepackage{extsizes}
\usepackage[super,sort&compress,comma]{natbib} 
\usepackage[version=3]{mhchem}
\usepackage[left=1.5cm, right=1.5cm, top=1.785cm, bottom=2.0cm]{geometry}
\usepackage{balance}
\usepackage{mathptmx}
\usepackage{sectsty}
\usepackage{graphicx} 
\usepackage{amsmath} 
\usepackage{ulem}
\usepackage{siunitx}
\DeclareSIUnit{\Molar}{\textsc{m}}
\usepackage{lastpage}
\usepackage[format=plain,justification=justified,singlelinecheck=false,font={stretch=1.125,small,sf},labelfont=bf,labelsep=space]{caption}
\usepackage{float}
\usepackage{fancyhdr}
\usepackage{fnpos}
\usepackage[english]{babel}
\addto{\captionsenglish}{%
  
}
\usepackage{array}
\usepackage{droidsans}
\usepackage{charter}
\usepackage[T1]{fontenc}
\usepackage[usenames,dvipsnames]{xcolor}
\usepackage{setspace}
\usepackage[compact]{titlesec}
\usepackage{hyperref}
\usepackage{color}
\hypersetup{pdftex, colorlinks=false,pdfborder={0 0 0}}
\usepackage{epstopdf}%This line makes .eps figures into .pdf - please comment out if not required.

\definecolor{cream}{RGB}{222,217,201}

\begin{document}

\pagestyle{fancy}
\thispagestyle{plain}
\fancypagestyle{plain}{
%%%HEADER%%%
\renewcommand{\headrulewidth}{0pt}
}
%%%END OF HEADER%%%

%%%PAGE SETUP - Please do not change any commands within this section%%%
\makeFNbottom
\makeatletter
\renewcommand\LARGE{\@setfontsize\LARGE{15pt}{17}}
\renewcommand\Large{\@setfontsize\Large{12pt}{14}}
\renewcommand\large{\@setfontsize\large{10pt}{12}}
\renewcommand\footnotesize{\@setfontsize\footnotesize{7pt}{10}}
\makeatother

\renewcommand{\thefootnote}{\fnsymbol{footnote}}
\renewcommand\footnoterule{\vspace*{1pt}% 
\color{cream}\hrule width 3.5in height 0.4pt \color{black}\vspace*{5pt}} 
\setcounter{secnumdepth}{5}

\makeatletter 
\renewcommand\@biblabel[1]{#1}            
\renewcommand\@makefntext[1]% 
{\noindent\makebox[0pt][r]{\@thefnmark\,}#1}
\makeatother 
\renewcommand{\figurename}{\small{Fig.}}
\sectionfont{\sffamily\Large}
\subsectionfont{\normalsize}
\subsubsectionfont{\bf}
\setstretch{1.125} %In particular, please do not alter this line.
\setlength{\skip\footins}{0.8cm}
\setlength{\footnotesep}{0.25cm}
\setlength{\jot}{10pt}
\titlespacing*{\section}{0pt}{4pt}{4pt}
\titlespacing*{\subsection}{0pt}{15pt}{1pt}
%%%END OF PAGE SETUP%%%

%%%FOOTER%%%
\fancyfoot{}
\fancyfoot[LO,RE]{\vspace{-7.1pt}\includegraphics[height=9pt]{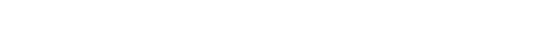}}
\fancyfoot[CO]{\vspace{-7.1pt}\hspace{13.2cm}\includegraphics{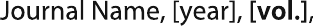}}
\fancyfoot[CE]{\vspace{-7.2pt}\hspace{-14.2cm}\includegraphics{head_foot/RF}}
\fancyfoot[RO]{\footnotesize{\sffamily{1--\pageref{LastPage} ~\textbar  \hspace{2pt}\thepage}}}
\fancyfoot[LE]{\footnotesize{\sffamily{\thepage~\textbar\hspace{3.45cm} 1--\pageref{LastPage}}}}
\fancyhead{}
\renewcommand{\headrulewidth}{0pt} 
\renewcommand{\footrulewidth}{0pt}
\setlength{\arrayrulewidth}{1pt}
\setlength{\columnsep}{6.5mm}
\setlength\bibsep{1pt}
%%%END OF FOOTER%%%

%%%FIGURE SETUP - please do not change any commands within this section%%%
\makeatletter 
\newlength{\figrulesep} 
\setlength{\figrulesep}{0.5\textfloatsep} 

\newcommand{\topfigrule}{\vspace*{-1pt}% 
\noindent{\color{cream}\rule[-\figrulesep]{\columnwidth}{1.5pt}} }

\newcommand{\botfigrule}{\vspace*{-2pt}% 
\noindent{\color{cream}\rule[\figrulesep]{\columnwidth}{1.5pt}} }

\newcommand{\dblfigrule}{\vspace*{-1pt}% 
\noindent{\color{cream}\rule[-\figrulesep]{\textwidth}{1.5pt}} }

\makeatother
%%%END OF FIGURE SETUP%%%

%%%TITLE, AUTHORS AND ABSTRACT%%%
\twocolumn[
  \begin{@twocolumnfalse}
{\includegraphics[height=30pt]{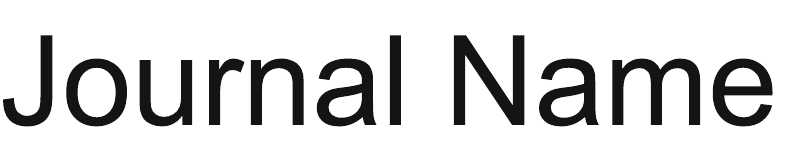}\hfill\raisebox{0pt}[0pt][0pt]{\includegraphics[height=55pt]{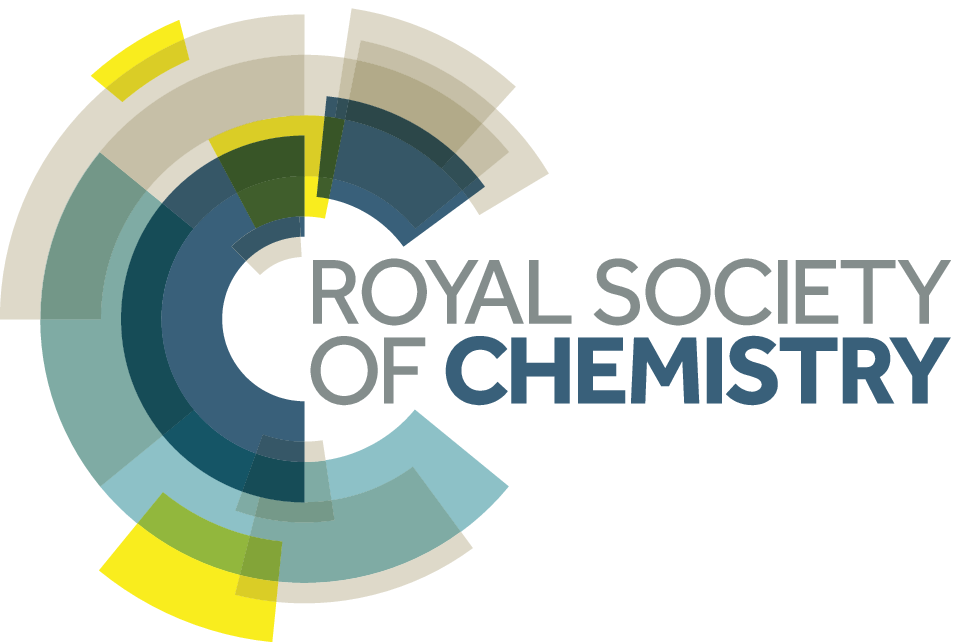}}\\[1ex]
\includegraphics[width=18.5cm]{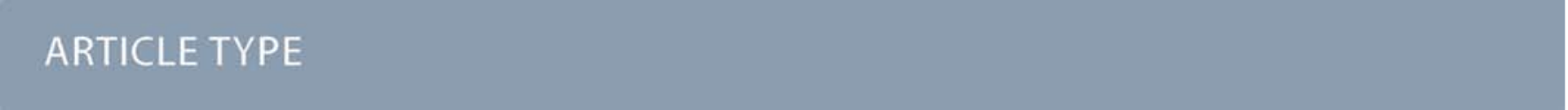}}\par
\vspace{1em}
\sffamily
\begin{tabular}{m{4.5cm} p{13.5cm} }

\includegraphics{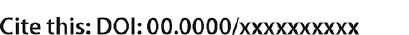} & \noindent\LARGE{\textbf{Designing Refractive Index Fluids using the Kramers--Kronig Relations$^\dag$}} \\
\vspace{0.3cm} & \vspace{0.3cm} \\

 & \noindent\large{Tianqi Sai,\textit{$^{a,b}$} Matthias Saba,\textit{$^{a}$} Eric R. Dufresne,\textit{$^{b}$} Ullrich Steiner,\textit{$^{\ast a}$} and Bodo D. Wilts\textit{$^{\ast a}$}} \\%Author names go here instead of "Full name", etc.

\includegraphics{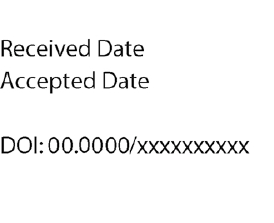} & \noindent\normalsize{For a number of optical applications, it is advantageous to precisely tune the refractive index of a liquid. Here, we harness a well-established concept in optics for this purpose. The Kramers-Kronig relation provides physical connection between the spectral variation of the (real) refractive index and the absorption coefficient. In particular a sharp spectral variation of the absorption coefficient gives rise to either an enhancement or reduction of the refractive index in the spectral vicinity of this variation. By using bright commodity dyes that fulfil this absorption requirement, we demonstrate the use of the Kramers-Kronig relation to predictively dial-in refractive index values in water solutions that are otherwise only attained by toxic specialised liquids.} \\

\end{tabular}

 \end{@twocolumnfalse} \vspace{0.6cm}

  ]
%%%END OF TITLE, AUTHORS AND ABSTRACT%%%

%%%FONT SETUP - please do not change any commands within this section
\renewcommand*\rmdefault{bch}\normalfont\upshape
\rmfamily
\section*{}
\vspace{-1cm}

%%%FOOTNOTES%%%

\footnotetext{\textit{$^{a}$~Adolphe Merkle Institute, University of Fribourg, Chemin des Verdiers 4, CH-1700 Fribourg, Switzerland.  E-mail: ullrich.steiner@unifr.ch; bodo.wilts@unifr.ch }}
\footnotetext{\textit{$^{b}$~Department of Materials, ETH Z\"urich, Vladimir-Prelog-Weg 5, CH-8093 Z\"urich, Switzerland. }}

%Please use \dag to cite the ESI in the main text of the article.
%If you article does not have ESI please remove the the \dag symbol from the title and the footnotetext below.
\footnotetext{\dag~Electronic Supplementary Information (ESI) available: Materials and Methods, Table S1 and Figs. S1-S4. See DOI: 00.0000/00000000.}
%additional addresses can be cited as above using the lower-case letters, c, d, e... If all authors are from the same address, no letter is required

% \footnotetext{\ddag~Additional footnotes to the title and authors can be included \textit{e.g.}\ `Present address:' or `These authors contributed equally to this work' as above using the symbols: \ddag, \textsection, and \P. Please place the appropriate symbol next to the author's name and include a \texttt{\textbackslash footnotetext} entry in the the correct place in the list.}

%%%END OF FOOTNOTES%%%

%%%MAIN TEXT%%%%
\section{Introduction}
The refractive index of a dielectric optical medium is typically considered to be an intrinsic property that is closely linked to the dipole moments of the atoms and molecules that constitute the optical medium \cite{hecht2016optics}. The tuning of optical substances -- liquid as well as solid -- is common place by either doping the optical medium with high dipole-strength substances or by designing the chemical nature of the medium.  The latter is for example exploited in the manufacture of liquids with high refractive indices that are used to increase the resolution in photo-lithography. Commercial immersion liquids with refractive indices beyond 1.60 are usually composed of methylene iodide, arsenic tribromide, arsenic disulfide, sulfur and selenium \cite{butler1933immersion,cargille2008immersion,darneal1948immersion,meyrowitz1951immersion,west1936immersion}. While widely used in industrial applications, these immersion liquids are toxic and the rational design of novel liquids based on non-toxic materials would be useful. This is in particular relevant for biological and medical high-resolution imaging, such as total internal reflection fluorescence microscopy \cite{Guo2018}, where high-index optics based on sapphire lenses facilitate super-resolution \cite{Laskar2016}.

To this end, it is useful to consider the fundamental theories of the propagation of light.  Based on the finite speed of the propagation of light, the consequences of causality on mathematical relationships of optical constants was considered nearly 100 years ago: For the scattering of light from an object, causality requires that ``no scattered wave can exist before the incident wave has reached the scattering center'' \cite{lucarini2005kramers}. Based on this consideration, Kramers \cite{Kramers} showed that the refractive index of a medium can be calculated from its absorption spectrum. Combined with Kronig's \cite{Kronig} argument that a ``dispersion relation is a sufficient and a necessary condition for strict causality to hold'' this establishes the well-known Kramers-Kronig relations, which constitute the connection between the in- and out-of-phase responses of a system to sinusoidal perturbations. Mathematically, the Kramers–Kronig relations connect the real and imaginary parts of any complex function that is analytic in the upper half-plane and vanishes sufficiently fast for large arguments \cite{lucarini2005kramers,nussenzveig1972causality}. Physically, such a function corresponds to an analytical linear response function of finite width in the time domain.

For an optical system with a complex refractive index $\mathbf{n}=n+i\kappa$, with $n(\omega)$ the `real' refractive index and $\kappa(\omega)$ the extinction coefficient, both of which vary with the angular frequency of the electromagnetic field $\omega$, the Kramers–Kronig relation can be written as \cite{lucarini2005kramers}
\begin{equation}
   n(\omega)-1=\frac{2}{\pi}\;\mathcal{P}\int_0^\infty \frac{\omega'\kappa(\omega')}{\omega'^2-\omega^2}\mathrm{d}\omega',
    \label{eq:KKomega}
\end{equation}
where $\mathcal{P}$ denotes the Cauchy principal value of the integral.

When used in spectroscopy, it is useful to write eqn (\ref{eq:KKomega}) in terms of the optical wavelength $\lambda = 2\pi c/\omega$, where $c$ is the speed of light. A substitution of the integration variable yields
\begin{equation}
 \Delta n(\lambda)=n(\lambda)-1=\frac{2}{\pi}\;\mathcal{P}\int_0^\infty \frac{\kappa(\lambda')}{\lambda'\left(1-\left(\frac{\lambda'}{\lambda}\right)^2\right)}\mathrm{d}\lambda'.
    \label{eq:KKlambda}
\end{equation}

An additionally requirement of the Kramers-Kronig relation is the superconvergence theorem \cite{lucarini2005kramers}, leading to
% \begin{equation}
%  \int_0^\infty \Delta n(\omega')\mathrm{d}\omega'=\lim_{\omega\to\infty}\left[\frac{\pi}{2}\omega\kappa(\omega)\right]=0,
%     \label{eq:asymomega}
% \end{equation}
% or 
\begin{equation}
 \int_0^\infty \frac{\Delta n(\lambda')}{\lambda'^2}\mathrm{d}\lambda'=\lim_{\lambda\to0}\left[\frac{\pi}{2}\frac{\kappa(\lambda)}{\lambda}\right]=0,
    \label{eq:asymlambda}
\end{equation}
defining the asymptotic limit of the extinction coefficient $\kappa$, which is related to the experimentally measured absorption coefficient by
\begin{equation}
    \alpha= \frac{4\pi}{\lambda}\kappa(\lambda).
 \label{eq:kappa}
\end{equation}

The use of Kramers-Kronig relation in optics is extremely well established \cite{lucarini2005kramers,faber2004oxygen,stavenga2013quantifying,stavenga2014oil}. It is often used to determine the spectral variation $n(\lambda)$ of an analyte from a measured absorption spectrum, and vice versa.  

While identical in terms of the underlying physics, this work explores an alternative approach to create new high refractive index liquids with a designed optical response, by exploring the Kramers-Kronig relation.  This concept was illustrated by  the refractive index dispersion of pigments in biological samples \cite{stavenga2013quantifying,stavenga2014oil}, the oxygen saturation levels of human blood \cite{faber2004oxygen}, and the high refractive index materials in the wing scales of Pierid butterflies \cite{wilts2017extreme}. 

\section{Results and discussion}
\subsection{Conceptualisation of the Kramers-Kronig approach}
\begin{figure}
    \centering
    \includegraphics[width=\linewidth]{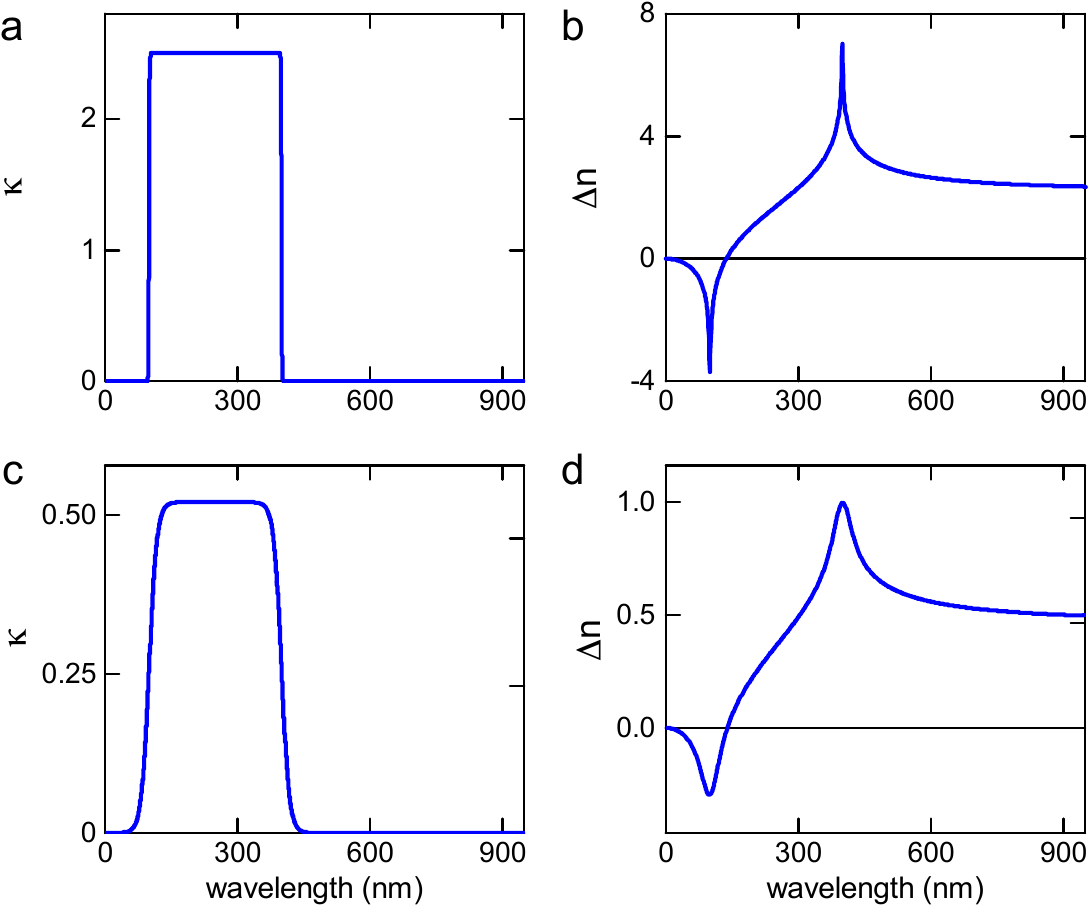}
    \caption{Refractive index spectra, calculated by numerically solving eqn (\ref{eq:KKlambda}). (a) Hypothetical extinction spectrum confined to the 100--400\,nm band.  (b) Refractive index change $\Delta n$ corresponding to (a). (c) Alternative hypothetical extinction spectrum and (d) corresponding $\Delta n$-spectrum. Similar plots with extinction spectra given by single Lorentzian peaks are shown in Fig.\ S3.}
    \label{fig:KK}
\end{figure}

Our concept is illustrated in Fig.\ \ref{fig:KK}. To explore the  physical limits of the Kramers-Kronig effect, i.e.\ the relation between absorption and refractive index change, we assume a hypothetical pigment with a sharply defined boxcar-shaped absorption band of $\kappa\,{=}\,2.5$ in the 100--\SI{400}{\nano\metre} wavelength range (Fig.\ \ref{fig:KK}a). Fig.\ \ref{fig:KK}b shows that in this hypothetical case, $\Delta n$ is very strongly enhanced: $\Delta n(\SI{400}{\nano\metre})\,{\approx}\, 7$; $\Delta n(\SI{400}{\nano\metre})\,{-}\,\Delta n(\infty)\,{\approx}\, 4.8$. A characteristic spectral feature of $\Delta n$ is the long wavelength decay, the region of ``normal'' dispersion, which can be approximated by the Cauchy equation \cite{jenkins1976fundamentals}
\begin{equation}
    n_\mathrm{C} = A + B/\lambda^2,
    \label{eq:Cauchy}
\end{equation}
where $A\,{=}\,(2/\pi)\int d\lambda'\,\kappa(\lambda')/\lambda'$ and $B\,{=}\,(2/\pi)\int d\lambda'\,\lambda'\kappa(\lambda')$ can be used as fit parameters. Note that each spectral feature in the the absorption spectrum gives rise to a long-wavelength Cauchy decay, which are additive \cite{lucarini2005kramers}, ultimately determining the overall refractive index variation of a given material in each transparent wavelength window.

While perhaps somewhat unrealistic, Fig.\ \ref{fig:KK}a,b illustrate the scope of our approach. Importantly, the enhancement of $\Delta n$ extends to wavelengths where the absorption coefficient is practically zero. The strong enhancement close to, but above, the absorption band makes this idea particularly attractive in applications where only narrow-band illumination is required, e.g.\ telecommunication \cite{agrawal2000nonlinear} or lenses for immersion lithography \cite{owa2004advantage}. By carefully selecting a short-wavelength absorber, a high refractive index can be dialled-in in a controlled fashion.

Fig.\ \ref{fig:KK}c,d shows a prediction which is closer to physical parameters that can be achieved for liquids: an extinction coefficient $\kappa\,{=}\,0.52$ which corresponds to $\alpha\,{\approx}\,\SI{10.3}{\per\micro\metre}$ at $\lambda\,{=}\,\SI{630}{\nano\metre}$ of a \SI{0.6}{\Molar} Brilliant Blue solution in water (see below). In this simulation, the absorption edge was smeared-out over ca.~\SI{35}{\nano\metre}. Here, an overall refractive index enhancement of ${\approx}\,1$ is expected at \SI{400}{\nano\metre} compared to the dye-free solution and a value of $\Delta n\,{\approx}\,0.8$ at \SI{430}{\nano\metre}, where the dye absorption has decayed to just a few percent of its maximal value. While perhaps not as impressive as the hypothetical case in Fig.\ \ref{fig:KK}b, this concept has the potential to enhance the refractive index of common liquids into the range commonly reserved to specialised toxic liquids \cite{cargille2008immersion}.

\subsection{Food dyes for water-based refractive index fluids}
To demonstrate the concept of refractive index tuning through the selection of an appropriate absorber, four commonly used non-toxic, water-soluble dyes were investigated: the triarylmethane dye ``Brilliant Blue'', the azo dye ``Allura Red'', the sulfonated  ``Quinoline Yellow'', and the arylsulfonate ``Pyranine''. The chemical structure and IUPAC names are given in Fig.\ S1 in the Supplementary Information (SI). The first four dyes are common food colorants \cite{martins2016food}, while the last is a fluorescent coloring agent used in hair dye and text markers. All four are water-soluble, non-toxic commodity compounds.

\begin{figure*}
    \centering
    \includegraphics[width=0.8\linewidth]{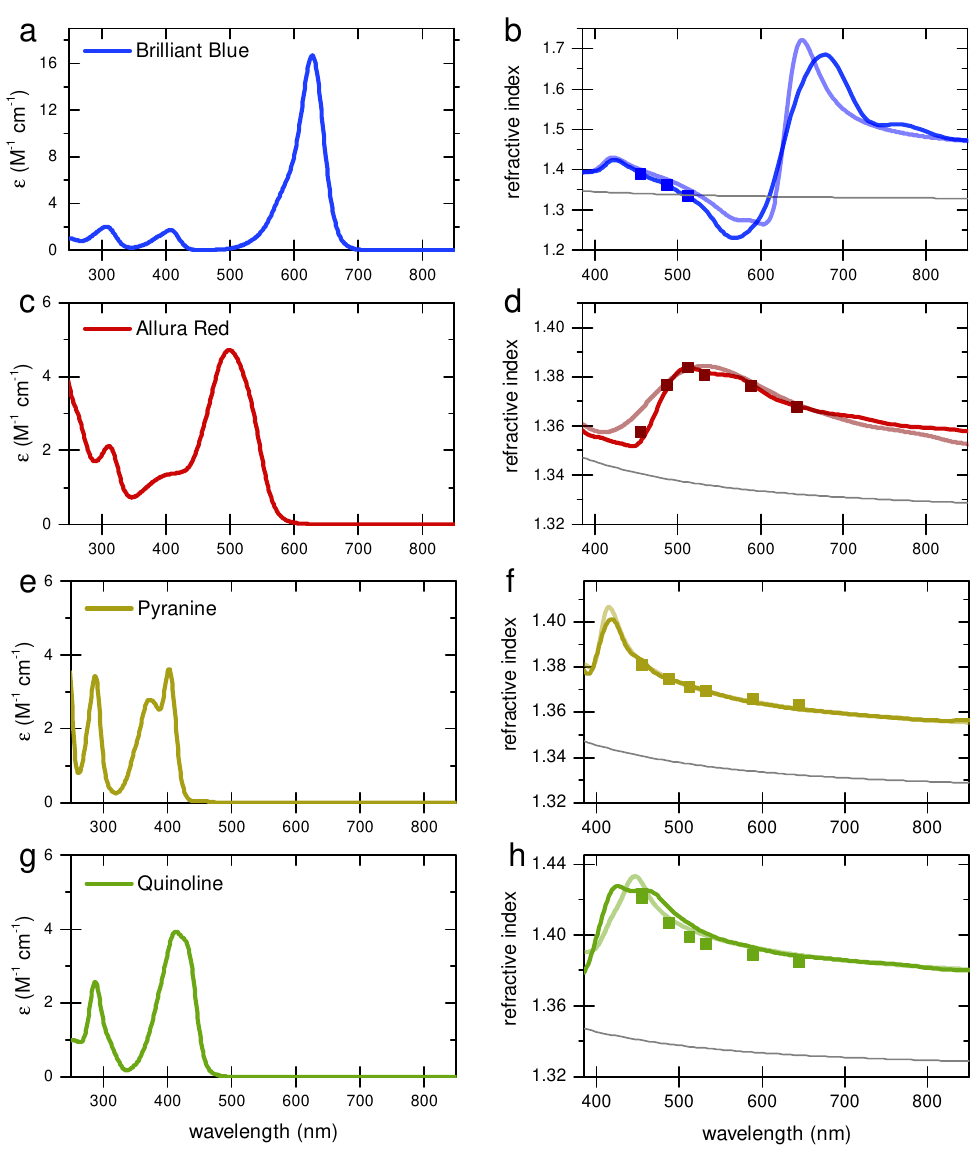}
    \caption{Molar absorption spectra $\varepsilon(\lambda)=\alpha(\lambda)/c_\mathrm{d}$ (left column) and refractive index variation (right column) of the four dye solutions in water, (a,b) \SI{0.6}{\Molar} Brilliant Blue, (c,d) \SI{0.3}{\Molar} Allura Red, (e,f) \SI{0.3}{\Molar} Pyranine, and (g,h) \SI{0.3}{\Molar} Quinoline Yellow. In the right column, the solid lines and the square symbols are ellipsometry and refractometry measurements, respectively. The light coloured lines are calculated using the Kramers-Kronig relations. The refractive index variation of water (grey lines) is shown for comparison.}
    \label{fig:allpigments}
\end{figure*}

To start our investigation, UV-Vis transmission spectra of all four dyes were acquired. Water solutions of the four dyes were prepared with series of concentrations $c_\mathrm{d}$ up to the solubility limit of the dyes (between \SI{300}{\milli\Molar} and \SI{600}{\milli\Molar}). The spectral absorption coefficients were then determined according to the Beer-Lambert law, shown in Fig.\ \ref{fig:allpigments}. The absorption coefficient of Allura Red shows rather broad spectral peaks with a broad decay toward long wavelengths and a peak absorption at about \SI{500}{\nano\metre}. In contrast, Brilliant Blue, Quinoline Yellow, and Pyranine show sharper spectral decays towards long wavelength at \SI{630}{\nano\metre}, \SI{412}{\nano\metre}, and \SI{428}{\nano\metre}, respectively, which combine the prerequisite for a strong $\Delta n$ enhancement at these wavelengths according to eqn (\ref{eq:KKlambda}) with the possibility to access a significant $\Delta n$-enhancement in the spectral domain where light absorption is insignificant (see Fig.\ \ref{fig:KK}). Particularly Brilliant Blue is an excellent candidate dye with an absorption coefficient of $17.25\,{\pm}\,\SI{0.04}{\per\Molar\per\micro\metre}$ at \SI{630}{\nano\metre}.

Spectroscopic ellipsometry was employed to measure the spectral variation $n(\lambda)$ of the various dye solutions, according to the protocol by Synowicki \textit{et al}. \cite{synowicki2004fluid}. In this method, dye solution drops are placed onto a rough glass substrate. The rough glass substrate enables the spreading of the water-dye solutions to provide a planar liquid surface and it suppresses the specular back reflection of the incident light beam from the liquid-substrate interface. Since only the phase relationship of the light reflected from the liquid-air surface is analysed, the ellipsometry data analysis is not very sensitive to the imaginary part of the refractive index $\kappa$. Therefore, only the real part of the refractive index $n(\lambda)$ was extracted from the ellipsometric data. To test the reliability of this approach, the variation of the refractive index of water was determined in the 400--\SI{1000}{\nano\metre} wavelength range and an agreement of better than 0.5\% with literature data was found (Fig.\ S2 in the SI). As a further test of the measurement protocol, a 6-wavelength (455--\SI{655}{\nano\metre}) Abbe-refractometer was also used, with a measurement accuracy of refractive index of $4\times 10^{-4}$.

The results of the ellipsometry and refractometry measurements for the highest concentration of the four dyes in water are shown in Fig.\ \ref{fig:allpigments}. A comparison with the spectral variation of the absorption curves bear out the qualitative variation of $\Delta n$ of Fig.\ \ref{fig:KK}, i.e.\ all curves exhibit the signature predicted by the Kramers-Kronig relations. 

To calculate the variation of $n(\lambda)$ from the absorption curves of Fig.\ \ref{fig:allpigments}, the limitations of the integration of eqn (\ref{eq:KKlambda}) have to be discussed. The prediction of the Kramers-Kronig relation is only precise if the variation of the extinction coefficient $\kappa$ is known for the entire spectral wavelength range ($0\to\infty)$. Since this is generally impossible, we exploit the limiting properties of eqn (\ref{eq:KKlambda}) given by the Cauchy eqn (\ref{eq:Cauchy}).

Considering, for example, an absorption spectrum with two distinct spectral features (Fig.\ S3 in the SI), the integral in eqn (\ref{eq:KKlambda}) can be decomposed into two separate integrations, the results of which are then added to yield the overall spectral response $n(\lambda)$ \cite{lucarini2005kramers}. In the range of the long-wavelength absorption peak, the $\Delta n$-contribution of the short-wavelength absorption peak can be approximated by eqn (\ref{eq:Cauchy}), which can be added to the result of eqn (\ref{eq:KKlambda}).

In the absence of spectral information below $\lambda = 250$\,nm, which strongly affects the long-wavelength variation of $\Delta n$, the following approach was therefore adopted.
\begin{enumerate}
    \item Calculation of $\kappa(\lambda)$ from the absorption curves in Fig.\ \ref{fig:allpigments}, using eqn (\ref{eq:kappa}), setting $\kappa=0$ for $\lambda<250$\,nm and $\lambda>1000$\,nm.
    \item Integration of eqn (\ref{eq:KKlambda}) to obtain $\Delta n(\lambda)$.
    \item Approximation of the refractive index variation of water $n_\mathrm{Cw}$ (Fig.\ S2 in the SI) by eqn (\ref{eq:Cauchy}).
    \item Calculation of $n_\mathrm{KK}(\lambda)\,{=}\,\Delta n(\lambda){+}n_\mathrm{Cw}(\lambda){+}n_\mathrm{Cd}(\lambda)$, with another Cauchy-type refractive index contribution of the dye, $n_\mathrm{Cd}(\lambda)$, as given by eqn (\ref{eq:Cauchy}). The coefficients $A$ and $B$ were fitted so that $n_\mathrm{KK}$ overlaps with the ellipsometry results in the short and long wavelength ranges far from the oscillatory feature in $\Delta n$ (Table~S1 in the SI).
\end{enumerate}

The motivation for the fitting procedure resulting in $n_\mathrm{Cd}$ stems from the fact that $\kappa(\lambda)$ of the dye is unknown for $\lambda<250$\,nm, the presence of which will give rise to a Cauchy-type decay for $\lambda>250$\,nm. Using this approach, an excellent agreement between calculated and measured refractive index spectra in the visible wavelength range was found (Fig.\ \ref{fig:allpigments}b,d,f,h).

\begin{figure}
    \centering
    \includegraphics[width=0.8\columnwidth]{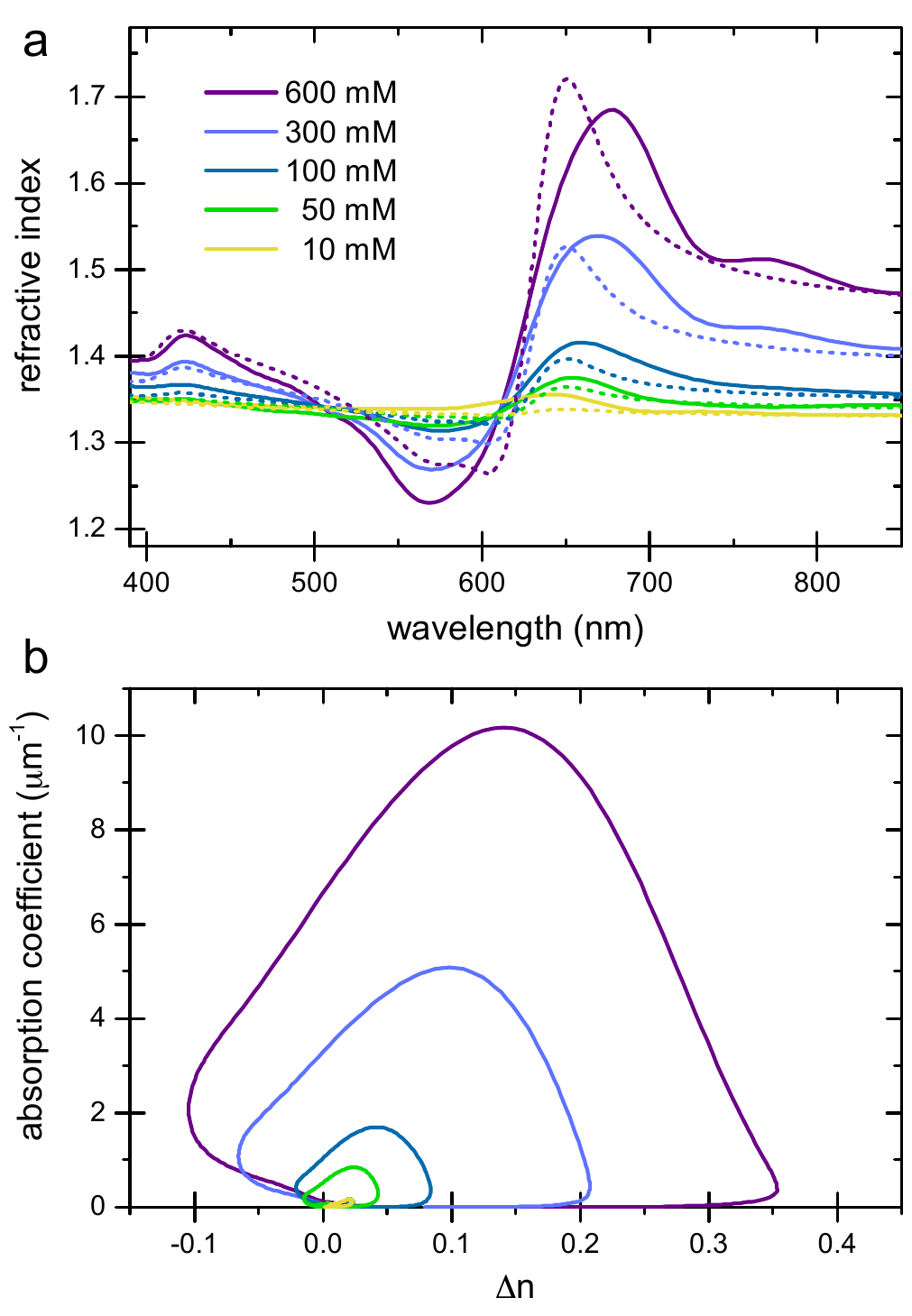}
    \caption{Refractive index variation of Brilliant Blue solutions in water as a function of dye concentration. (a) Ellipsometry measurements (solid lines) are compared to $n_\mathrm{KK}(\lambda)$ calculated using the Kramers-Kronig relation (eqn (\ref{eq:KKlambda}), dashed lines). Note the increase in the refractive index at short wavelengths, which is indicative of additional dye absorption peaks below \SI{250}{\nano\metre} (see text) \cite{BB_note}. (b) A plot of the absorption coefficient vs.\ $\Delta n$ clearly reveals a substantial $\Delta n$-enhancement in spectral regions where the absorption coefficient is practically zero.}  
    \label{fig:BB_c_dependence}
\end{figure}

\subsection{Refractive index tuning}
The refractive index of the liquid can be further tuned by changing the concentration of the dye. Fig.\ \ref{fig:BB_c_dependence}a demonstrates that is it possible to precisely dial-in a desired refractive index at a target wavelength, by adjusting the concentration of the Brilliant Blue dye. Fig.\ \ref{fig:BB_c_dependence}b clearly demonstrates that $\Delta n$-enhancements of up to $\approx0.35$ can be achieved in spectral regions where the absorption coefficient is close to zero. The concentration series in Fig.\ \ref{fig:BB_c_dependence}a further gives evidence for the presence of a finite $n_\mathrm{Cd}$ in the visible. With increasing dye concentration, not only does the oscillatory feature of the refractive index become increasingly pronounced, stemming from the single absorption peak at \SI{630}{\nano\metre}, the ``baseline'' at 380--\SI{400}{\nano\metre} also increases with dye concentration. Fig.\ \ref{fig:KK} illustrates that $\Delta n\,{\to}\,0$ for shorter wavelengths compared to the absorption peak. Therefore, the baseline variation indicates the presence of a Cauchy term, eqn (\ref{eq:Cauchy}), stemming from dye absorption peaks below \SI{380}{\nano\metre}.

The comparison of the calculated $n_\mathrm{KK}$ with the measured ellipsometry spectra is very good, given the fact that spectral information outside the 250--1000\,nm wavelength window is missing and Cauchy approximations were employed instead. Note also that while the fit determining $n_\mathrm{Cd}(\lambda)$ improves the Kramers-Kronig description of the experimental data, the predictive quality of $n_\mathrm{KK}$ even in the absence of any fitting (i.e.\ setting $n_\mathrm{Cd}=0$) is very good, showing the strength of this approach (Fig.\ S4 in the SI). %The Cauchy fit parameters used to determine $n_\mathrm{Cd}$ Fig.\ \ref{fig:allpigments} are given in Table~\ref{tab:Cauchytable} in the SI. 

\begin{figure}
    \centering
    \includegraphics[width=\columnwidth]{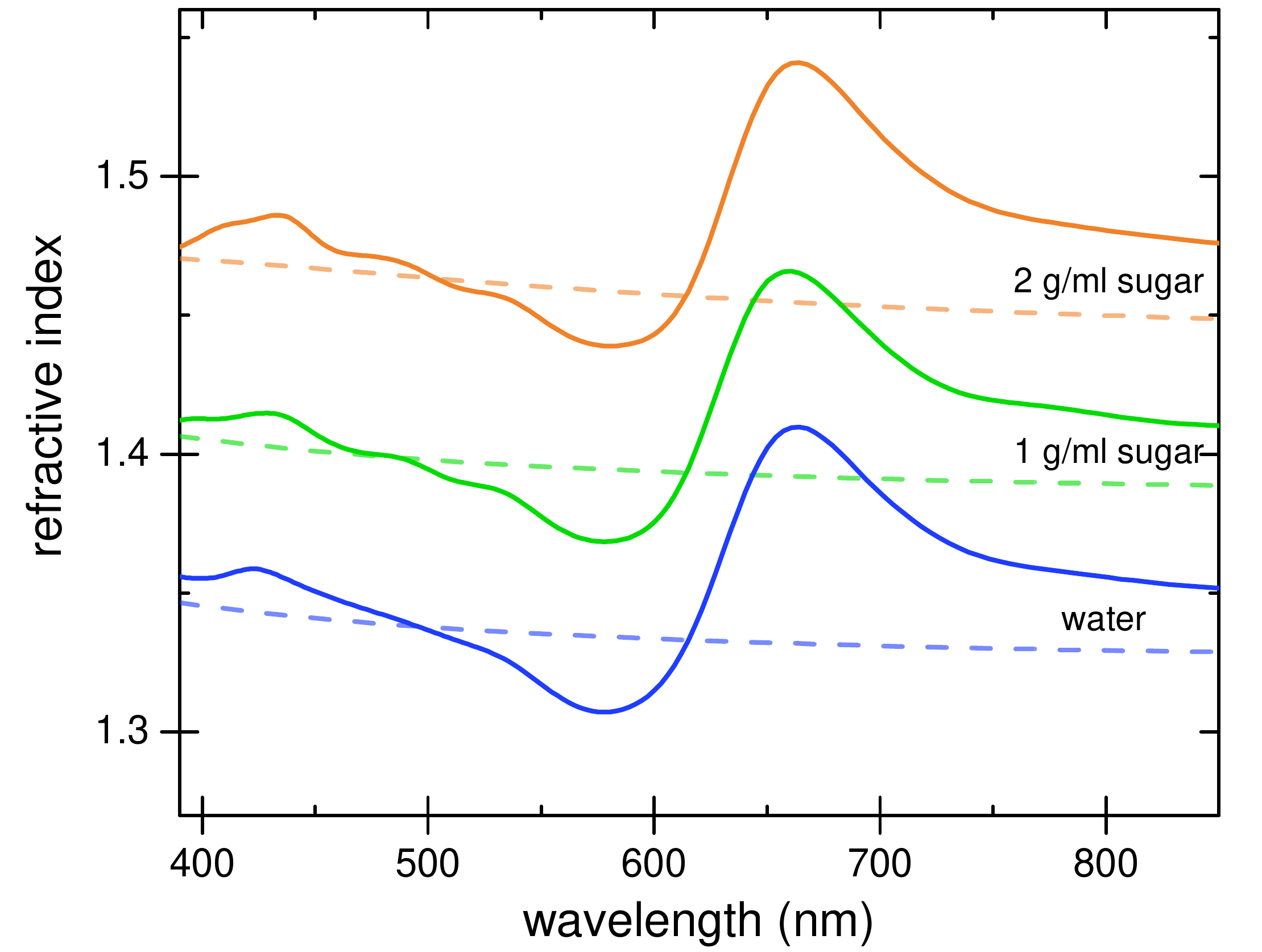}
    \caption{Refractive index variation of 100\,mM Brilliant Blue dissolved in in water (bottom curve), 1\,g/ml sugar solution (middle curve) and 2\,g/ml sugar solution (top curve). The dashed lines show the dye-free reference.}
    \label{fig:BB_sugar}
\end{figure}

A further way to control the refractive index of a liquid at a given refractive index is the combination of the ``traditional'' method, the addition of a transparent additive, with the present Kramers-Kronig approach. Fig.\ \ref{fig:BB_sugar} shows the measured refractive index spectra of 0.1\,M Brilliant Blue dissolved in two sugar solutions. The combination of raising the Cauchy background (eqn (\ref{eq:Cauchy})) with the Kramers-Kronig response of the dye provides a versatile way to control the refractive index of the solution across large parts of the optical spectrum.

\section{Conclusions}
In summary, we have harnessed a nearly 100-year old concept in optics for the design of refractive indices in liquids. When employing narrow bandwidth illumination, as e.g.\ provided by lasers, the well-established Kramers-Kronig relations that link the refractive index with the absorption coefficient provide an approach to specifically dial-in a target refractive index at any narrow wavelength band. Using common food dyes, refractive indices close to 1.7 were obtained. It should be possible to further enhance the refractive index variation by developing dyes that optimise the  Kramers-Kronig effect, featuring high absorption coefficients with a sharp long-wavelength decay. Alternatively, it might be possible to suspend organic or inorganic nanoparticles in solution to harness the high extinction coefficients of crystalline materials \cite{palmer2020highly,wilts2017extreme}. Finally, while the refractive index enhancement at long wavelengths is larger than the refractive index reduction of the Kramers-Kronig oscillation caused by a well-defined absorption peak at shorter wavelengths, this effect can also be employed for the targeted refractive index reduction, e.g.\ to index match or produce narrow-band anti-reflective coatings.

\section*{Conflicts of interest}
There are no conflicts of interests to declare.

\section*{Acknowledgements}
This work was partially funded by the ERC Advanced Grant (H2020-Prismoid, to US) and through the Swiss National Science Foundation through the NCCR \textit{Bio-inspired Materials} (to TS, ERD, US, BDW) and the Ambizione programme (168223, to BDW). 
%%%END OF MAIN TEXT%%%


\balance


%%%REFERENCES%%%
\bibliography{bibliography} %You need to replace "rsc" on this line with the name of your .bib file

\begin{mcitethebibliography}{23}
\bibitem[BB_()]{BB_note}
Compared to the trend given by the lower dye concentrations, the experimental
  curve of the \SI{600}{\milli\Molar} Brilliant Blue solution has a lower than
  expected refractive index maximum. Since this concentration is close to the
  solubility limit, a small fraction of the dye may have separated from the
  solution cause an effective solution concentration below
  \SI{600}{\milli\Molar}. By extrapolation the probable solution concentration
  of this data set was \SI{540}{\milli\Molar}.\relax
\mciteBstWouldAddEndPunctfalse
\mciteSetBstMidEndSepPunct{\mcitedefaultmidpunct}
{}{\mcitedefaultseppunct}\relax
\EndOfBibitem
\end{mcitethebibliography}
\bibliographystyle{rsc} %the RSC's .bst file
\newpage
\setcounter{figure}{0}   
\setcounter{table}{0} 
\setcounter{equation}{0} 
\renewcommand{\thefigure}{S\arabic{figure}}
\renewcommand{\thetable}{S\arabic{table}}
\renewcommand{\theequation}{S\arabic{equation}}
% \renewcommand{\figurename}{\small{Fig.}~}
%%%TITLE, AUTHORS AND ABSTRACT%%%
\twocolumn[
  \begin{@twocolumnfalse}
% {\includegraphics[height=30pt]{head_foot/journal_name}\hfill\raisebox{0pt}[0pt][0pt]{\includegraphics[height=55pt]{head_foot/RSC_LOGO_CMYK}}\\[1ex]
% \includegraphics[width=18.5cm]{head_foot/header_bar}}\par
% \vspace{1em}
\sffamily
%\begin{tabular}{m{4.5cm} p{13.5cm} }

\noindent\LARGE{\textbf{Designing Refractive Index Fluids using the Kramers--Kronig Relations}} \\
\vspace{0.3cm} \\
\noindent\LARGE{-- Supplementary Information --}\\
\vspace{0.3cm} \\
\noindent\large{Tianqi Sai,\textit{$^{a,b}$} Matthias Saba,\textit{$^{a}$} Eric R. Dufresne,\textit{$^{b}$} Ullrich Steiner,\textit{$^{\ast a}$} and Bodo D. Wilts\textit{$^{\ast a}$}} \\%Author names go here instead of "Full name", etc.

\footnotesize{\textit{$^{a}$~Adolphe Merkle Institute, University of Fribourg, Chemin des Verdiers 4, CH-1700 Fribourg, Switzerland.  E-mail: ullrich.steiner@unifr.ch; bodo.wilts@unifr.ch }}\\
\footnotesize{\textit{$^{b}$~Department of Materials, ETH Z\"urich, Vladimir-Prelog-Weg 5, CH-8093 Z\"urich, Switzerland. }}
%\includegraphics{head_foot/dates} & \\
%\end{tabular}

\bigskip

 \end{@twocolumnfalse} \vspace{0.6cm}

  ]
%%%END OF TITLE, AUTHORS AND ABSTRACT%%%

%%%FONT SETUP - please do not change any commands within this section
\renewcommand*\rmdefault{bch}\normalfont\upshape
\rmfamily
\section*{}
\vspace{-1cm}

%%%FOOTNOTES%%%

% \footnotetext{\textit{$^{a}$~Adolphe Merkle Institute, University of Fribourg, Chemin des Verdiers 4, CH-1700 Fribourg, Switzerland.  E-mail: ullrich.steiner@unifr.ch; bodo.wilts@unifr.ch }}
% \footnotetext{\textit{$^{b}$~Department of Materials, ETH Z\"urich, Vladimir-Prelog-Weg 5, CH-8093 Z\"urich, Switzerland. }}
%\onecolumn

\section*{Materials and methods}
\subsection*{Materials}
\begin{figure*}
    \centering
    \includegraphics[width=0.8\linewidth]{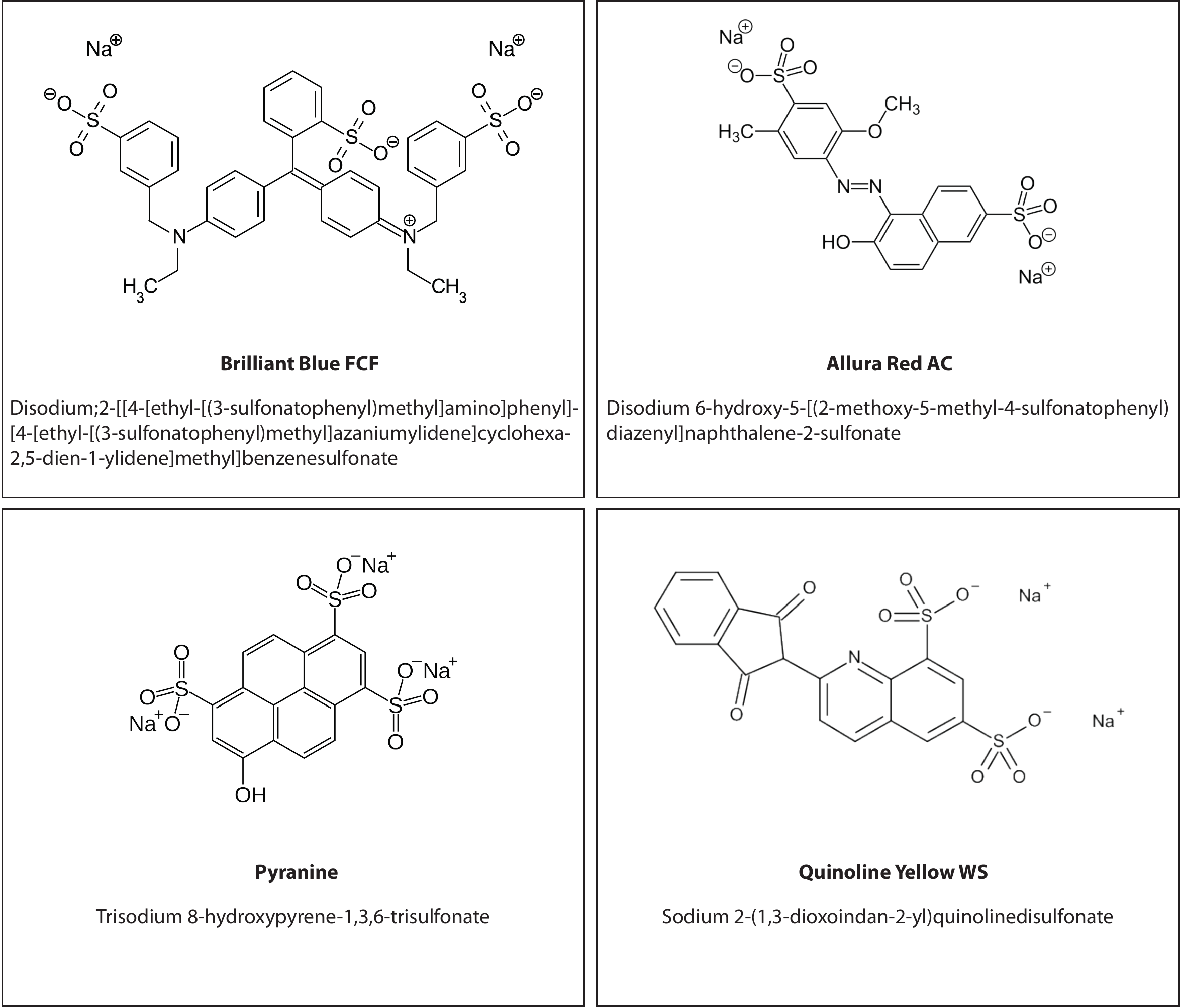}
    \caption{Chemical structures and IUPAC names of the dyes used in this study}
    \label{fig:Dyes}
\end{figure*}

The four dyes Brilliant Blue, Allura Red, Pyranine and Quinoline Yellow were purchased from fastcolours.com. Their chemical structures and IUPAC full names are given in Fig.\ \ref{fig:Dyes}. Stock solutions with concentrations of 600\,mM for Brilliant Blue and 300\,mM for  the other dyes were prepared and subsequently diluted to lower concentrations. The absorption of those dyes in the visible range is mainly due to the resonance among the benzene rings and double bonds in the structures\footnotemark{}.
%\cite{gurses2016classification}.

\subsection*{Optical characterization}
Transmittance spectra were recorded using a spectrometer (Maya Pro, Ocean Optics) with a wavelength range from 250 -- 1000\,nm. The dye solutions were filled into a quartz cuvette with a 1\,cm light path and then placed into a Thorlabs CVH100 cuvette holder, which has two fiber adapters on opposite sides. Light from a halogen/deuterium or a xenon light source (DHL-Bal or HPX-2000, both Ocean Optics) passes through the sample and is guided into the spectrometer via optical fibres with a \SI{200}{\micro\metre} diameter (QP200-2-SR-BX, Ocean Optics). A water-filled cuvette was used as reference.

Assuming a small extinction coefficient $\kappa\ll n$, the absorption coefficient $\alpha$ is experimentally measured through application of the Beer-Lambert law
\begin{equation}
    \frac{I(\lambda)}{I_0(\lambda)} = e^{-i\alpha(\lambda) d},
    \label{eq:Beer}
\end{equation}
with $I$, $I_0$  the transmitted intensities of light after passing through a dye solution of thickness $d$ and a dye-free reference, respectively. The molar absorption coefficient is then defined as %\cite{BB_note2}
\begin{equation}
    \varepsilon(\lambda) = \frac{\alpha(\lambda)}{c_\mathrm{d}},
\end{equation}
where $c_\mathrm{d}$ is the molar concentration of the dye. The relation to the extinction coefficient $\kappa$ is given by eqn (\ref{eq:kappa}).
% \begin{equation}
%     \alpha= \frac{4\pi}{\lambda}\kappa(\lambda).
%  \label{eq:kappa}
% \end{equation}

%Absorbance $A(\lambda)$ spectra were determined from transmittance %spectra $T(\lambda)$ using the  Beer-Lambert law
%\begin{equation}
%    A(\lambda)=-\log %T(\lambda)=-\log\left[\frac{I(\lambda)}{I_0(\lambda)}\right],
%    \label{eq:BeerLambert}
%\end{equation}
%where $I$ and $I_0$ are the transmitted and incident light intensities, %respectively.  The molar extinction coefficient is given by
%\begin{equation}
%    \varepsilon(\lambda)=2.3 A(\lambda)/ (c_\mathrm{d} d),
%\end{equation}
%where $c_\mathrm{d}$ is the dye concentration and d is the light path %length through the cuvette. 

\subsection*{Refractive index measurements}
The refractive index dispersion was measured with an optical ellipsometer (alpha-se, JA Wollam). For this, the liquid was deposited onto rough, frosted glass slides (Marienfeld) and the dielectric function was extracted as described previously by Synowicki \textit{et al.}\cite{synowicki2004fluid}.

The refractive index of the fluids was additionally assessed with an automatic refractometer (Abbemat MW, Anton-Paar) at six discrete wavelengths (455.0\,nm, 488.0\,nm, 513.3\,nm, 533.1\,nm, 589.3\,nm and 644.2\,nm), with a RI range of 1.30 - 1.72 nD and an accuracy of $\pm 0.00004$\,nD.

%\section*{Cauchy parameters}

\begin{table}[h]
    \centering
    \begin{tabular}{c c c }
	    \hline 
	    dye & A & B [$\si{\nano\metre}^2$] \\ 
	    \hline \hline
	    Brilliant Blue & 0.064 $\pm$ 0.008 & 0  \\ 
%	    \hline 
	    Allura Red & 0.062 $\pm$ 0.004  & 511 $\pm$ 25 \\ 
%	    \hline 
	    Pyranine & 0.0143 $\pm$ 0.001 & 946 $\pm$ 31 \\ 
%	    \hline 
	    Quinoline Yellow & 0.0393 $\pm$ 0.003 & 2258 $\pm$ 69 \\ 
	    \hline 
    \end{tabular} 
    \caption{Fit parameters $A$, $B$, determining $n_\mathrm{Cd}(\lambda)$ according to eqn (\ref{eq:Cauchy}) for the four  dyes.}
    \label{tab:Cauchytable}
\end{table}

\footnotetext{A. G\"urses, M. A{\c{c}}{\i}ky{\i}ld{\i}z, K. G\"une{\c{s}}, K\"ubra and M. S. G\"urses, Dyes and Pigments, Springer, 2016, pp.\ 31–45.}
\pagebreak

\clearpage
\begin{figure}[H]
    \centering
    \includegraphics[width=\linewidth]{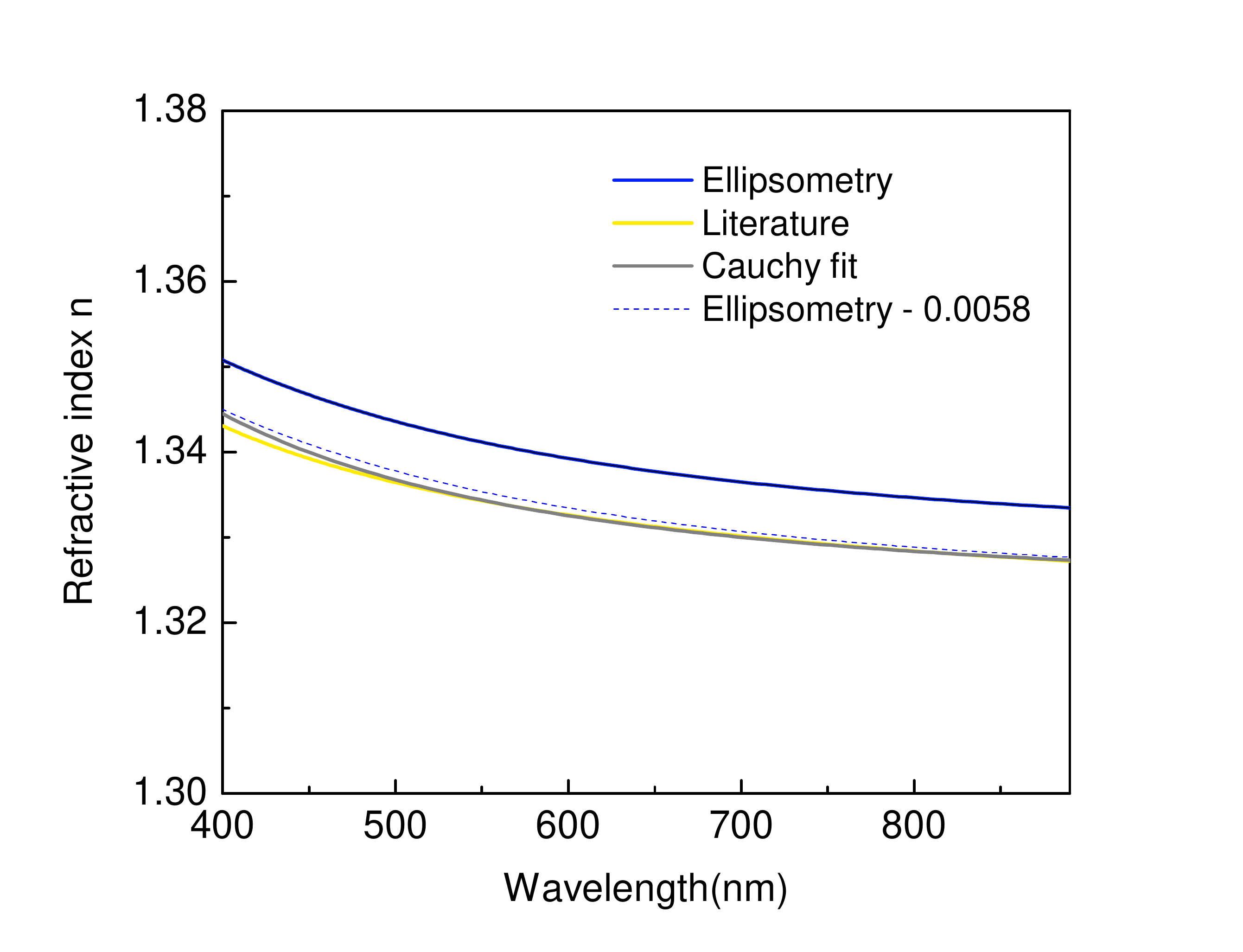}
    \caption{Ellipsometry measurement of the refractive index of water, in comparison with literature values. In this study, a Cauchy fit (eqn (\ref{eq:Cauchy})) was used to account for the spectral variation of the refractive index of water.  Despite the close agreement of the dispersion curve from literature and the ellipsometry measurements, an offset of 0.0058 was deducted from all ellipsometry measurements to account for the discrepancy found in this measurement.}
    \label{fig:Water}
\end{figure}

\begin{figure}[H]
    \centering
    \includegraphics[width=\columnwidth]{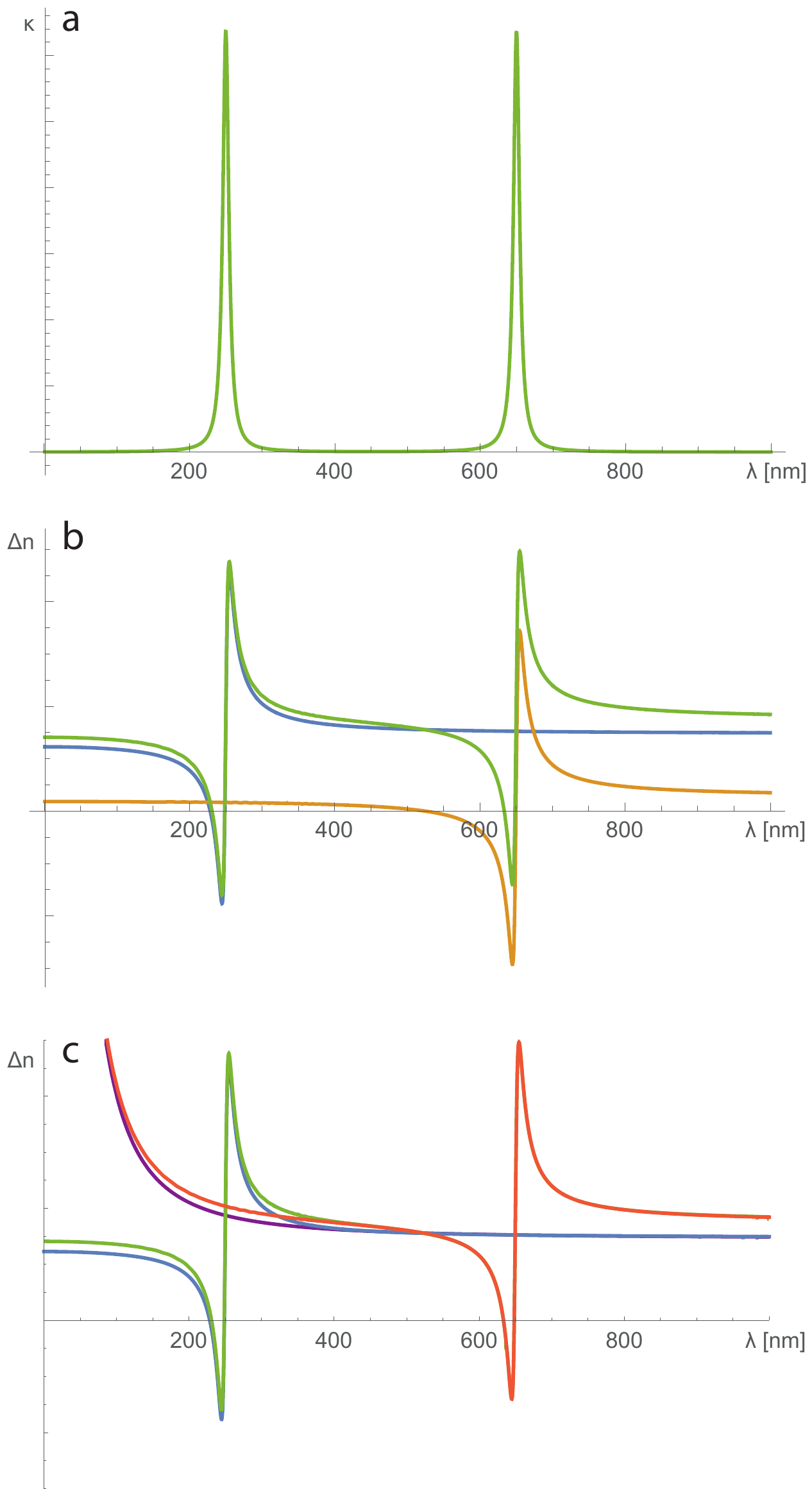}
    \caption{Model calculations illustrating the additivity of the results of eqn (\ref{eq:KKlambda}).  (a) Hypothetical extinction spectrum consisting of two Lorentz distributions centred at 250\,nm and and 650\,nm.  (b) The integration of eqn (\ref{eq:KKlambda}) of extinction spectra containing each peak separately result in the orange and blue lines, which were added to yield the green spectrum. This resulting green spectrum is indistinguishable from the one stemming from an integration employing the double-peak spectrum. In (c) the Cauchy equation eqn (\ref{eq:Cauchy}) was fitted to the blue spectrum for wavelengths greater than 400\,nm, resulting in the purple line. The addition of this Cauchy spectrum with the orange line in (b) (resulting from the Kramers-Kronig integration of only the right peak) gives rise to the red line in (c), which approximates the exact spectrum (green line) very closely for wavelengths above 400\,nm.}
    \label{fig:Cauchy}
\end{figure}

\clearpage

\begin{figure*}
    \centering
    \includegraphics[width=0.8\linewidth]{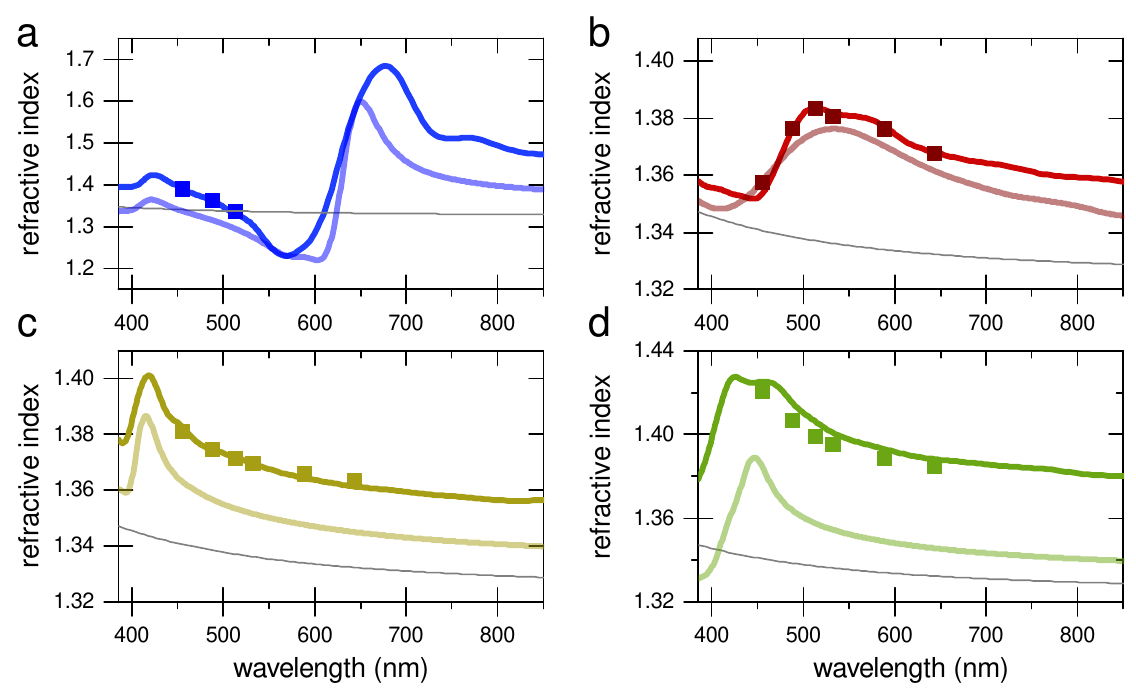}
    \caption{Refractive index data from Fig.\ \ref{fig:allpigments} (solid lines) and the predictions of Kramers-Kronig relation $n_\mathrm{KK}$ (light coloured lines). In the calculation of $n_\mathrm{KK}$ only the Cauchy correction for water $n_\mathrm{Cw}$ was taken into account and the final Cauchy fit was omitted ($n_\mathrm{Cd}=0$, see text).}
    \label{fig:noCorr}
\end{figure*}

\end{document}